\documentclass[aps,prl,twocolumn,superscriptaddress,letterpaper]{revtex4}
\usepackage{mathrsfs}
\usepackage{amsmath}
\usepackage{bm}
\usepackage{color}
\usepackage{graphicx}
\usepackage{hyperref}
\usepackage{pdfpages}
\usepackage{pgffor}

\begin{document}


\title{Efficient Framework for Solving Plasma Waves with Arbitrary Distributions}

\author{Huasheng Xie}
\email[]{Email: huashengxie@gmail.com, xiehuasheng@enn.cn} 
\affiliation{Hebei Key Laboratory of Compact Fusion, Langfang 065001, China}
\affiliation{ENN Science and Technology Development Co., Ltd., Langfang 065001, China}

\date{\today}

\begin{abstract}
Plasma, which constitutes 99\% of the visible matter in the universe, is characterized by a wide range of waves and instabilities that play a pivotal role in space physics, astrophysics, laser-plasma interactions, fusion research, and laboratory experiments. The linear physics of these phenomena is described by kinetic dispersion relations (KDR). However, solving KDRs for arbitrary velocity distributions remains a significant challenge, particularly for non-Maxwellian distributions frequently observed in various plasma environments. This work introduces a novel, efficient, and unified numerical framework to address this challenge. The proposed method rapidly and accurately yields all significant solutions of KDRs for nearly arbitrary velocity distributions, supporting both unstable and damped modes across all frequencies and wavevectors. The approach expands plasma species' velocity distribution functions using a series of carefully chosen orthogonal basis functions and employs a highly accurate rational approximation to transform the problem into an equivalent matrix eigenvalue problem, eliminating the need for initial guesses. The efficiency and versatility of this framework are demonstrated, enabling simplified studies of plasma waves with arbitrary distributions. This advancement paves the way for uncovering new physics in natural plasma environments, such as spacecraft observations in space plasmas, and applications like wave heating in fusion research.
\end{abstract}


\maketitle

The vast majority—99\%—of the visible matter in the universe exists in the plasma state, which is characterized by charged particles interacting with electromagnetic fields. These interactions give rise to numerous waves and instabilities. Since the 1950s, kinetic theory has provided a framework to describe these phenomena \cite{Stix1992}, the linear physics of which are well understood through dispersion relations (DR). However, solving the theoretical models is challenging due to mathematical complexities such as singularities and infinite double integrals. Analytical solutions are only feasible for limited cases, and numerical methods have primarily focused on specific distributions (e.g., Maxwellian).

\begin{figure*}
\centering
\includegraphics[width=17cm]{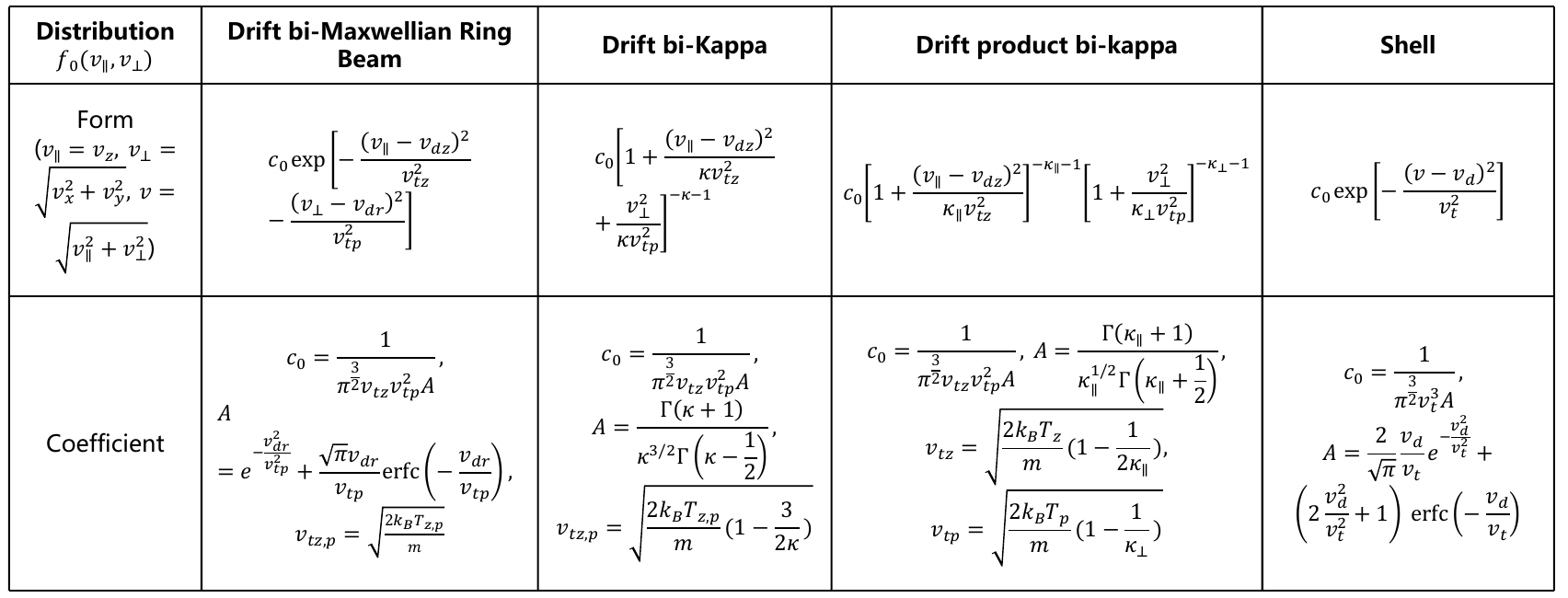}\\
\caption{Typical non-Maxwellian velocity distributions encountered in space and laboratory plasmas.}\label{fig:fveq}
\end{figure*}

An efficient numerical solver for plasma waves could significantly advance the field, akin to how traditional approaches revolutionized condensed matter physics \cite{Car1985} or how artificial intelligence (AI) transformed protein structure prediction \cite{Jumper2021}. Non-Maxwellian distributions, frequently observed in both space and laboratory plasmas, demand an effective and general solution framework (several typical distributions are illustrated in Figure \ref{fig:fveq}). Since the 1990s, several attempts have been made to address this challenge \cite{Matsuda1992}, with notable progress in recent years \cite{Hellinger2011, Astfalk2017, Verscharen2018, Irvine2018}. However, the high computational cost (e.g., hours of CPU time) and the reliance on initial guesses to locate suitable solutions have limited the widespread use of these tools for practical analyses. While faster solvers are available for Maxwellian-based \cite{Ronnmark1982, Xie2016, Xie2019} and kappa distributions \cite{Astfalk2015, Lopez2021, Bai2025}, a general solver for arbitrary distributions remains elusive.

In this work, we introduce a groundbreaking numerical method that, for the first time, meets nearly all five key requirements for an ideal kinetic dispersion relation (KDR) solver: (1) Fast; (2) Accurate; (3) Capable of resolving damped modes; (4) Able to reliably find solutions or obtain all solutions; (5) Compatible with arbitrary distributions. Notably, the fourth requirement is achieved seamlessly in conjunction with the fifth, representing a major breakthrough. We demonstrate in this work how these challenges are resolved with this novel approach.

Though a wide range of scenarios can be addressed, we focus on the electromagnetic magnetized model, limiting our discussion to linear plasma waves in an infinite, uniform, and homogeneous system. We assume a background magnetic field ${\bm B}_0=(0,0,B_0)$ and a wave vector ${\bm k}=(k_x,0,k_z)=(k\sin\theta,0,k\cos\theta)$, where $k_\parallel=k_z$ and $k_\perp=k_x$. The system consists of $S$ species, indexed by $s=1, 2, \cdots, S$, with each species characterized by an electric charge $q_s$, mass $m_s$, and density $n_{s0}$. The dispersion relation is given by \cite{Summers1994,Gurnett2005,Xie2019}:
\begin{eqnarray}
  \bar{D}(\omega,{\bm k})
  =\big|{\bm K}(\omega,{\bm k})+({\bm k}{\bm k}-k^2{\bm I})\frac{c^2}{\omega^2}\big|=0,
\end{eqnarray}
where ${\bm K}={\bm I}+{\bm Q}={\bm I}-\frac{{\bm \sigma}}{i\omega\epsilon_0},~~{\bm Q}=-\frac{{\bm \sigma}}{i\omega\epsilon_0}$, and
\begin{eqnarray}\label{eq:sigma}
  {\bm \sigma} =-i\sum_s\frac{q_s^2n_{s0}}{m_s}\sum_{n=-\infty}^{\infty}\int_{-\infty}^{\infty}\int_0^{\infty}\frac{2\pi v_\perp dv_\perp dv_\parallel}{\omega-n\omega_{cs}-k_\parallel v_\parallel}{\bm \Pi}_s,
\end{eqnarray}
with
\begin{eqnarray}
{\bm \Pi}_s=\begin{bmatrix}
  A_s\frac{n^2v_\perp}{\mu_s^2}J_n^2 & iA_s\frac{nv_\perp}{\mu_s}J_nJ'_n & B_s\frac{nv_\perp}{\mu_s}J_n^2\\
  -iA_s\frac{nv_\perp}{\mu_s}J_nJ'_n & A_sv_\perp {J'_n}^2 & -iB_s v_\perp J_nJ'_n\\
  A_s\frac{nv_\parallel}{\mu_s}J_n^2 & iA_sv_\parallel J_nJ'_n & B_sv_\parallel J_n^2
\end{bmatrix},
\end{eqnarray}
where $\mu_s=\frac{k_\perp v_\perp}{\omega_{cs}}$, 
$A_s=\Big(1-\frac{k_\parallel v_\parallel}{\omega}\Big)\frac{\partial f_{s0}}{\partial v_\perp}+\frac{k_\parallel v_\perp}{\omega}\frac{\partial f_{s0}}{\partial v_\parallel}$, and 
$B_s=\frac{n\omega_{cs} v_\parallel}{\omega v_\perp}\frac{\partial f_{s0}}{\partial v_\perp}+\Big(1-\frac{n\omega_{cs}}{\omega}\Big)\frac{\partial f_{s0}}{\partial v_\parallel}$.  Eq.~(\ref{eq:sigma}) is valid for non-relativistic, arbitrary gyrotropic distributions. Here, $J_n=J_n(\mu_s)$ is the Bessel function of the first kind of order $n$, with its derivative $J'_n=dJ_n(\mu_s)/d\mu_s$. Additional terms are defined as ${\bm n}=\frac{{\bm k}c}{\omega}$, $\omega_{cs}=\frac{q_sB_0}{m_s}$, $\omega_{ps}=\sqrt{\frac{n_{s0}q_s^2}{\epsilon_0m_s}}$, and $c=\frac{1}{\sqrt{\mu_0\epsilon_0}}$, where $\omega_{ps}$ and $\omega_{cs}$ represent the plasma and cyclotron frequencies, respectively, $c$ is the speed of light, $\epsilon_0$ is the permittivity of free space, $\mu_0$ is the permeability of free space, and $\bm n$ is the refractive index vector.

We express the normalized distribution function of species $s$ as a series expansion in basis functions $ f_{s0}=\sum_{l=-\infty}^{\infty}\sum_{m=-\infty}^{\infty}a_{s0,lm}\rho_{sz,l}(v_\parallel)u_{sx,m}(v_\perp)$, where $-\infty<v_\parallel<\infty$ and $0\leq v_\perp<\infty$. Here, $\rho_{sz,l}$ and $u_{sx,m}$ are basis functions for the parallel and perpendicular directions, respectively, which decouple the velocity integrals in Eq.~(\ref{eq:sigma}). The choice of basis functions significantly impacts performance. Early attempts for the 1D case used orthogonal basis functions such as Hermite, Legendre, or Chebyshev polynomials \cite{Robinson1990}, as well as Fourier-based rational functions \cite{Weideman1994,Weideman1995,Xie2013} for the parallel integral. Other approaches, such as using fitted/interpolated functions for the parallel direction and Legendre basis functions for the perpendicular direction, have been applied to study wave heating in the ion cyclotron range of frequencies (ICRF) \cite{Brambilla2013,Bilato2012}.

Our goal is to solve the DR for arbitrary distributions accurately and to obtain all important solutions without requiring an initial guess. To achieve this, we expand the DR into a rational form of the wave frequency $\omega=\omega_r+i\omega_i$ \cite{Xie2016,Xie2019,Xie2024}, where $\omega_i>0$ corresponds to instability and $\omega_i<0$ corresponds to a damped mode. Direct integration on grids \cite{Matsuda1992,Hellinger2011,Verscharen2018} or spline function fitting \cite{Astfalk2017,Irvine2018} cannot yield a rational dispersion function with respect to $\omega$. 
We identify two types of expansions that meet our requirements. The first is the GPDF basis \cite{Xie2013}, which is related to Fourier bases and thus offers exponential convergence with high accuracy and efficiency. This approach can be transformed into a matrix method similar to the $\kappa$-distribution case \cite{Bai2023,Bai2025}. The second is the Hermite basis, which is closely related to the Maxwellian distribution. The Hermite basis yields a form suitable for J-pole expansion and facilitates a matrix approach \cite{Xie2016,Xie2019} with a smaller matrix dimension, making it possible to obtain all solutions without requiring an initial guess. In this work, we demonstrate results using the Hermite-Hermite (HH) basis for both parallel and perpendicular directions. Detailed derivations of the equations, along with results using GPDF-Hermite (GH) and GPDF-GPDF (GG) bases, are provided in the supplementary material.

For the HH expansion, we use the following form:
\begin{eqnarray}
f_{s0}(v_\parallel,v_\perp)=c_{s0}\sum_{l=0}^{\infty}\sum_{m=0}^{\infty}a_{s,lm}\cdot g_{sz,l}(v_\parallel)\cdot g_{sx,m}(v_\perp),
\end{eqnarray}
where $g_{sz,l}(v_\parallel) = \Big(\frac{v_\parallel-d_{sz}}{L_{sz}}\Big)^l e^{-\Big(\frac{v_\parallel-d_{sz}}{L_{sz}}\Big)^2}$, $g_{sx,m}(v_\perp) = \Big(\frac{v_\perp-d_{sx}}{L_{sx}}\Big)^m e^{-\Big(\frac{v_\perp-d_{sx}}{L_{sx}}\Big)^2}$. Here, $L_{sz}$ and $L_{sx}$ are the velocity widths, while $d_{sz}$ and $d_{sx}$ are the drift velocities. The normalization coefficient is given by $c_{s0}=\frac{1}{\pi^{3/2}L_{sz}L_{sx}^2R_s}$, where 
$R_s=\exp\Big(-\frac{d_{sx}^2}{L_{sx}^2}\Big)+\frac{\sqrt{\pi}d_{sx}}{L_{sx}}{\rm erfc}\Big(-\frac{d_{sx}}{L_{sx}}\Big)$, and ${\rm erfc}(-x) = 1 - {\rm erf}(-x) = 1 + {\rm erf}(x)$ is the complementary error function. The coefficients $a_{s,lm}$ can be readily calculated using orthogonal Hermite basis functions.

The benefits of this choice of basis functions include: (1) It naturally reduces to the drift bi-Maxwellian ring-beam case \cite{Xie2019} by retaining only the lowest-order term with $a_{s,00}=1\neq0$. (2) The parallel integral can be simplified to involve only a single Maxwellian $Z$ function, and the perpendicular integral can also be reduced to a single term initially. 
The second benefit allows the J-pole matrix method from BO \cite{Xie2019} to be directly applied here, with the matrix dimension remaining unchanged, though large values of $J$ may be required for accurate computation. Thus, a slight modification of the Maxwellian-based BO code can be used for generalized arbitrary distributions.

We define
$f_{s0,lm}(v_\parallel,v_\perp)\equiv a_{s,lm}g_{sz,l}(v_\parallel)g_{sx,m}(v_\perp)$,
$f_{s0z,l}(v_\parallel)\equiv g_{sz,l}(v_\parallel)$, $f_{s0x,m}(v_\perp)\equiv g_{sx,m}(v_\perp)$.
Hence, we have $\frac{\partial f_{s0z,l}(v_\parallel)}{\partial v_\parallel}=-\frac{1}{L_{sz}}\Big[2f_{s0z,l+1}-lf_{s0z,l-1}\Big]$, $\frac{\partial f_{s0x,m}(v_\perp)}{\partial v_\perp}=-\frac{1}{L_{sx}}\Big[2f_{s0x,m+1}-mf_{s0x,m-1}\Big]$.
Here, we set  $f_{s0z,l}=f_{s0x,m}=0$ for all $l,m<0$. The above derivatives equations are valid for all $l,m=0,1,2,\cdots$.

For parallel integral [${\rm Im}(\zeta_{sn})>0$], we define
\begin{eqnarray}\nonumber
Z_{l,p}(\zeta_{sn})\equiv-\frac{k_\parallel}{L_{sz}^p\sqrt{\pi}}\int_{-\infty}^{\infty}\frac{v_\parallel^p (\frac{v_\parallel-d_{sz}}{L_{sz}})^le^{-(\frac{v_\parallel-d_{sz}}{L_{sz}})^2}}{\omega-k_\parallel v_\parallel-n\omega_{cs}}dv_\parallel\\
=\frac{1}{\sqrt{\pi}}\int_{-\infty}^{\infty}\frac{g_{l,p}(z)}{z-\zeta_{sn}}dz,~~~g_{l,p}(z)\equiv(z+d)^pz^le^{-z^2}.
\end{eqnarray}
We have $Z_{0,0}(\zeta)=Z(\zeta)$ and redifine  $Z_{l,0}\equiv Z_l$ and $I_{l}\equiv\pi^{-1/2}\int_{-\infty}^{\infty}x^{l}e^{-x^2}dx$,
with $\zeta_{sn}=\frac{\omega-k_\parallel d_{sz}-n\omega_{cs}}{k_\parallel L_{sz}}$, $d=\frac{d_{sz}}{L_{sz}}$,  $p=0,1,2$. These plasma dispersion functions can be computed efficiently to high accuracy and are also analytically continuous for ${\rm Im}(\zeta_{sn}) \leq 0$ \cite{Xie2013}.
For the perpendicular integral, we define
{\tiny
\begin{eqnarray}\nonumber
\Gamma_{\{a,b,c\}n,m,p}(a_s,d_s)\equiv\frac{1}{L_{sx}^{p+1}}\int_0^{\infty}v_\perp^p\{J_n^2(\frac{k_\perp v_\perp}{\omega_{cs}}),J_n(\frac{k_\perp v_\perp}{\omega_{cs}})J'_n(\frac{k_\perp v_\perp}{\omega_{cs}}),\\\nonumber
 {J'_n}^2(\frac{k_\perp v_\perp}{\omega_{cs}})\}(\frac{v_\perp-d_{sx}}{L_{sx}})^me^{-(\frac{v_\perp-d_{sx}}{L_{sx}})^2} dv_\perp\\
=\int_0^{\infty}x^p\{J_n^2(a_sx),J_n(a_sx)J_n'(a_sx),{J'_n}^2(a_sx)\}(x-d_s)^me^{-(x-d_s)^2} dx,\end{eqnarray}}
with $a_{s}=k_\perp\rho_{cs}$,  $\rho_{cs}=L_{sx}/\omega_{cs}$, $d_s=d_{sx}/L_{sx}$, $p=0,1,2,3$.  These integrals can be computed numerically \cite{Xie2019}.

\begin{figure}
\centering
\includegraphics[width=8.5cm]{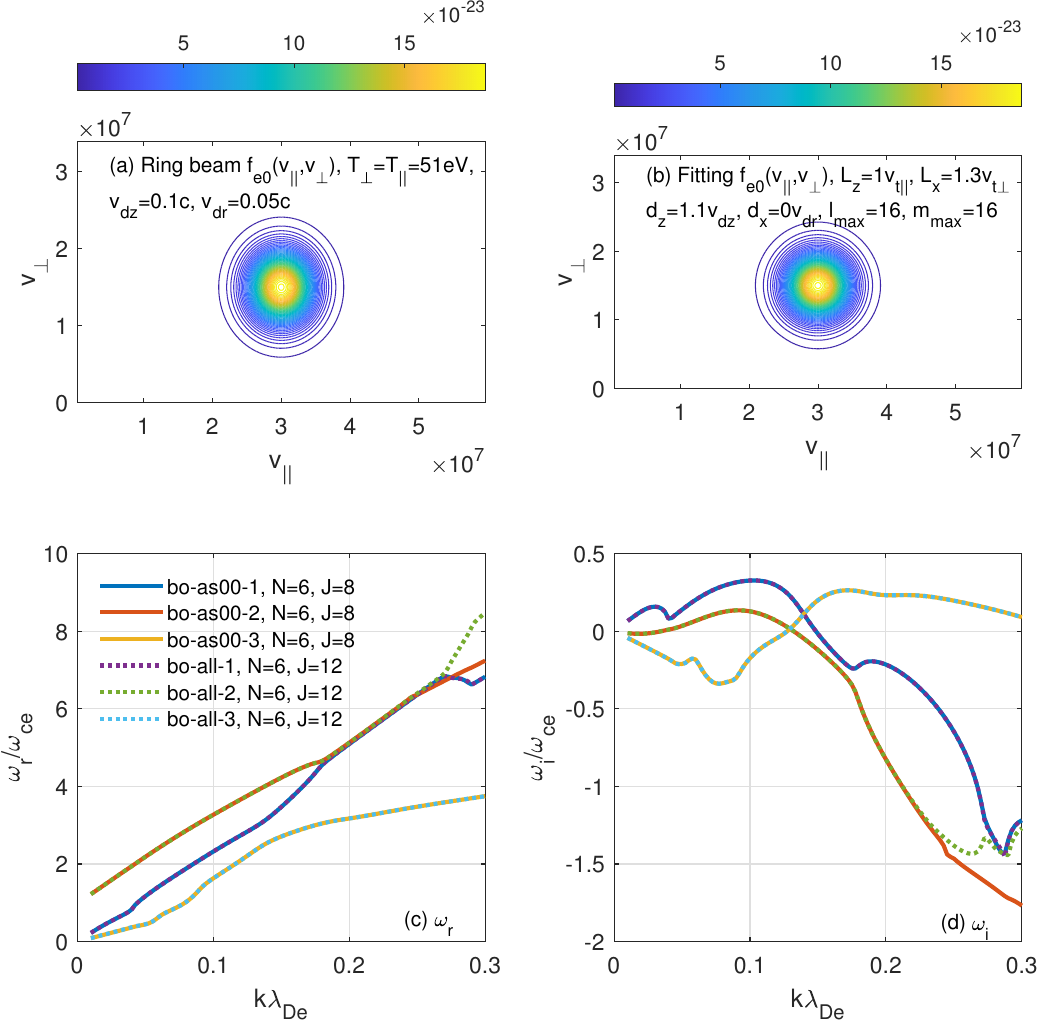}\\
\caption{Comparison of unstable wave solutions under a ring beam electron distribution at $\theta=40^\circ$, using both analytical and fitted distributions, reveals good agreement except in regions with strong damping. Here, "as00" represents the analytical distribution, and "all" denotes the use of Hermite expansion. Solving for 120 wave vector points and obtaining all solutions requires approximately 33 seconds for the analytical case (N=6, J=8), 64 seconds for the fitted case (N=6, J=8), and 200 seconds for the fitted case (N=6, J=12).}\label{fig:bo_Umeda12_ringbeam_theta=40}
\end{figure}

\begin{figure}
\centering
\includegraphics[width=8.5cm]{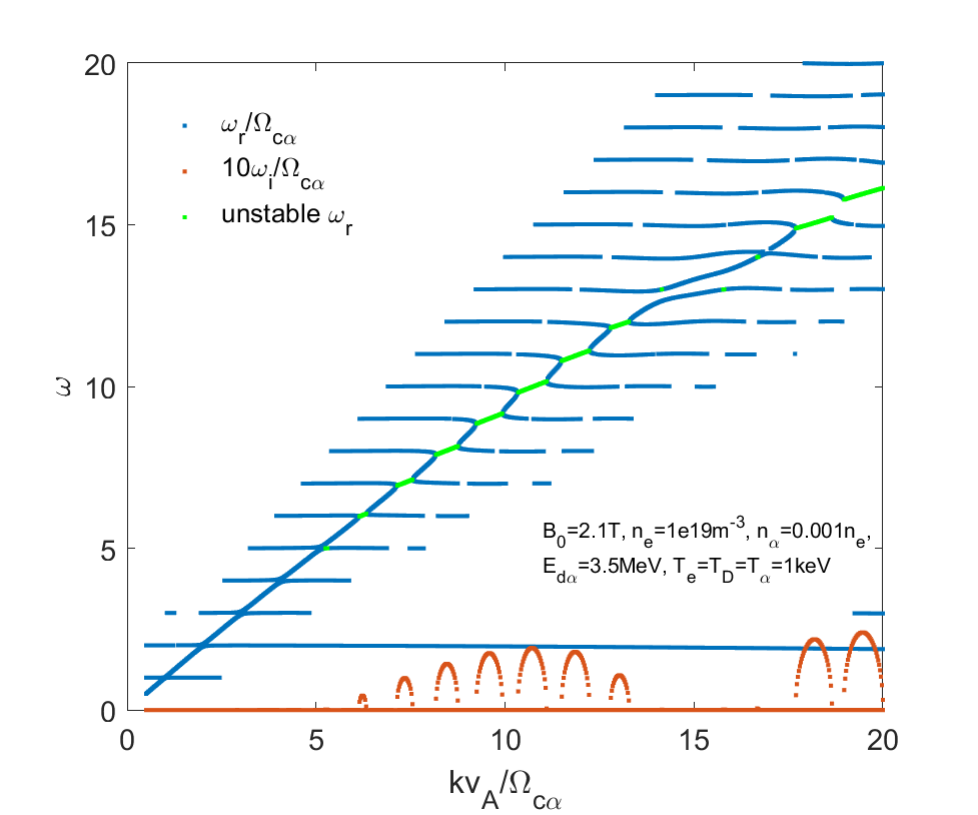}\\
\caption{Ion cyclotron emission in magnetically confined fusion plasma with drift ring ion beam distributions at $\theta=89.5^\circ$.}\label{fig:bo_Irvine18_ICE}
\end{figure}

For the Maxwellian $Z$ function J-pole expansion, we have\cite{Xie2019}
\begin{eqnarray}
  Z_l(\zeta)\simeq\sum_{j=1}^{J}\frac{b_jc_j^l}{\zeta-c_j}.
\end{eqnarray}
Here, we have used\cite{Xie2024} $\sum_jb_j=-1$, $\sum_jb_jc_j=0$, $\sum_jb_jc_j^2=-1/2$, $\sum_jb_jc_j^3=0$, $\cdots$. 
This also means that if we need a higher-order Hermite fitting of the distribution function, the corresponding coefficients should also be kept to a similar order. Typically, we need to set $J \geq l_{\text{max}} + 4$. To ensure double precision, we have calculated up to $J = 24$. Hence, the present solver can support $l_{\text{max}} \simeq 20$. While higher values of $l_{\text{max}}$ can also be computed, the accuracy would decrease.

To seek an equivalent linear system, Maxwell's equations are
\begin{subequations} \label{eq:em3dmaxw}
\begin{eqnarray}
  & \partial_t {\bm E} = c^2\nabla\times{\bm B}-{\bm J}/\epsilon_0,\\
  & \partial_t {\bm B} = -\nabla\times{\bm E},
\end{eqnarray}
\end{subequations}
which do not need to be changed. We only need to seek a new linear system for ${\bm J}=\bm{\sigma}\cdot{\bm E}$.

Considering the definition $\bm \sigma_s = -i \epsilon_0 \omega \bm Q_s = -i \epsilon_0 \frac{\omega_{ps}^2}{\omega} \bm P_s$, after the J-pole expansion, we have
$P_{s11}=\sum_{n} \frac{2}{R_s} \frac{n^2\omega_{cs}^2}{k_\perp^2L_{sx}^2}\sum_{l,m}a_{s,lm}\Big\{(\frac{n\omega_{cs}}{k_\parallel L_{sz}}Z_{l}-I_{l})(2\Gamma_{an,m+1,0}-m\Gamma_{an,m-1,0})+ \Gamma_{an,m,1}\frac{L_{sx}^2}{L_{sz}^2}(2Z_{l+1}-lZ_{l-1})\Big\}\simeq\sum_{n,j}\frac{2}{R_s} \frac{n^2\omega_{cs}^2}{k_\perp^2 L_{sx}^2}\frac{k_zL_{sz}b_j}{\omega-c_{snj}} \sum_{l,m}a_{s,lm}\Big\{\frac{n\omega_{cs}}{k_\parallel L_{sz}}c_j^{l}(2\Gamma_{an,m+1,0}-m\Gamma_{an,m-1,0})+ \Gamma_{an,m,1}\frac{L_{sx}^2}{L_{sz}^2}[2c_j^{l+1}-lc_j^{l-1}]\Big\}- {\frac{2}{R_s} \frac{\omega_{cs}^2}{k_\perp^2 L_{sx}^2}\sum_{l,m}a_{s,lm}\Big\{I_{l}\sum_{n}n^2(2\Gamma_{an,m+1,0}-m\Gamma_{an,m-1,0})\Big\}}=\sum_{n,j}\frac{p_{11snj}}{\omega-c_{snj}}-1$, where $p_{11snj}= \frac{2}{R_s} \frac{n^2\omega_{cs}^2}{k_\perp^2 L_{sx}^2}{k_zL_{sz}b_j} \sum_{l,m}a_{s,lm}\Big\{\frac{n\omega_{cs}}{k_\parallel L_{sz}}c_j^{l}(2\Gamma_{an,m+1,0}-m\Gamma_{an,m-1,0})+ \Gamma_{an,m,1}\frac{L_{sx}^2}{L_{sz}^2}[2c_j^{l+1}-lc_j^{l-1}]\Big\}$. Other terms are similar.

It is thus easy to find that after the J-pole expansion, the relations between $\bm J$ and $\bm E$ have the following form{\tiny
\begin{equation}\label{eq:JmE}
   \frac{{\bm \sigma}}{-i\epsilon_0}
    =\left( \begin{array}{ccc}
    \frac{b_{11}}{\omega}+\sum_{snj}\frac{b_{snj11}}{\omega-c_{snj}} & \frac{b_{12}}{\omega}+\sum_{snj}\frac{b_{snj12}}{\omega-c_{snj}}
    & \frac{b_{13}}{\omega}+\sum_{snj}\frac{b_{snj13}}{\omega-c_{snj}}\\
    \frac{b_{21}}{\omega}+\sum_{snj}\frac{b_{snj21}}{\omega-c_{snj}} & \frac{b_{22}}{\omega}+\sum_{snj}\frac{b_{snj22}}{\omega-c_{snj}}
    & \frac{b_{23}}{\omega}+\sum_{snj}\frac{b_{snj23}}{\omega-c_{snj}} \\
    \frac{b_{31}}{\omega}+\sum_{snj}\frac{b_{snj31}}{\omega-c_{snj}} & \frac{b_{32}}{\omega}+\sum_{snj}\frac{b_{snj32}}{\omega-c_{snj}}
    & \frac{b_{33}}{\omega}+\sum_{snj}\frac{b_{snj33}}{\omega-c_{snj}}
    \end{array}\right).
\end{equation}}
 
Combining Eqs.~(\ref{eq:em3dmaxw}) and (\ref{eq:JmE}), the equivalent linear system for the electromagnetic DR can be obtained as
\begin{equation}\label{eq:BOkem}
    \left\{ \begin{array}{ccc}
    \omega v_{snjx} &=& c_{snj} v_{snjx} + b_{snj11} E_x + b_{snj12} E_y + b_{snj13} E_z, \\
    \omega j_x &=&  b_{11} E_x + b_{12} E_y + b_{13} E_z, \\
    i J_x\epsilon_0 &=& j_x+\sum_{snj}v_{snjx}, \\
    \omega v_{snjy} &=& c_{snj} v_{snjy} + b_{snj21} E_x + b_{snj22} E_y + b_{snj23} E_z, \\
    \omega j_y &=&  b_{21} E_x + b_{22} E_y + b_{23} E_z, \\
    iJ_y/\epsilon_0 &=& j_y+\sum_{snj}v_{snjy}, \\
    \omega v_{snjz} &=& c_{snj} v_{snjz} + b_{snj31} E_x + b_{snj32} E_y + b_{snj33} E_z, \\
    \omega j_z &=&  b_{31} E_x + b_{32} E_y + b_{33} E_z, \\
    iJ_z/\epsilon_0 &=& j_z+\sum_{snj}v_{snjz}, \\
    \omega E_x &=& c^2k_z B_y-iJ_x/\epsilon_0, \\
    \omega E_y &=& -c^2k_z B_x +c^2k_x B_z-iJ_y/\epsilon_0,\\
    \omega E_z &=& -c^2k_x B_y-iJ_z/\epsilon_0, \\
    \omega B_x &=& -k_z E_y, \\
    \omega B_y &=& k_z E_x - k_x E_z, \\
    \omega B_z &=& k_x E_y,
    \end{array}\right.
\end{equation}
which yields a sparse matrix eigenvalue problem that can be readily solved using standard eigenvalue libraries. The symbols such as $v_{snjx}$, $j_{x,y,z}$, and $J_{x,y,z}$ do not have direct physical meanings but are analogous to the perturbed velocity and current density in the fluid-based derivations of plasma waves~\cite{Xie2014}. However, the elements of the eigenvector 
$(E_x, E_y, E_z, B_x, B_y, B_z)$ still represent the original perturbed electric and magnetic fields. Thus, the polarization of the solutions can also be obtained in a straightforward manner.
The dimension of the matrix is $N_N=3\times (N_{SNJ}+1)+6=3\times
\{[S\times(2\times N+1)]\times J+1\}+6$, where $N$ is the number of harmonics retained for magnetized species, and $J$ is the order of the J-pole expansion for the $Z$ function. A surprising and unexpected feature of the above HH expansion is that the matrix structure and dimension are identical to the Maxwellian distribution case. Thus, the computation time remains the same, except for the time required to calculate the initial matrix coefficients. Other existing solvers for arbitrary distribution KDRs (e.g.,~\cite{Verscharen2018, Irvine2018}) incur significantly higher computational costs compared to the Maxwellian case. Similar to Maxwellian-based cases~\cite{Xie2016, Xie2019}, the above approach avoids the singularity in the DR.

\begin{figure}
\centering
\includegraphics[width=8.5cm]{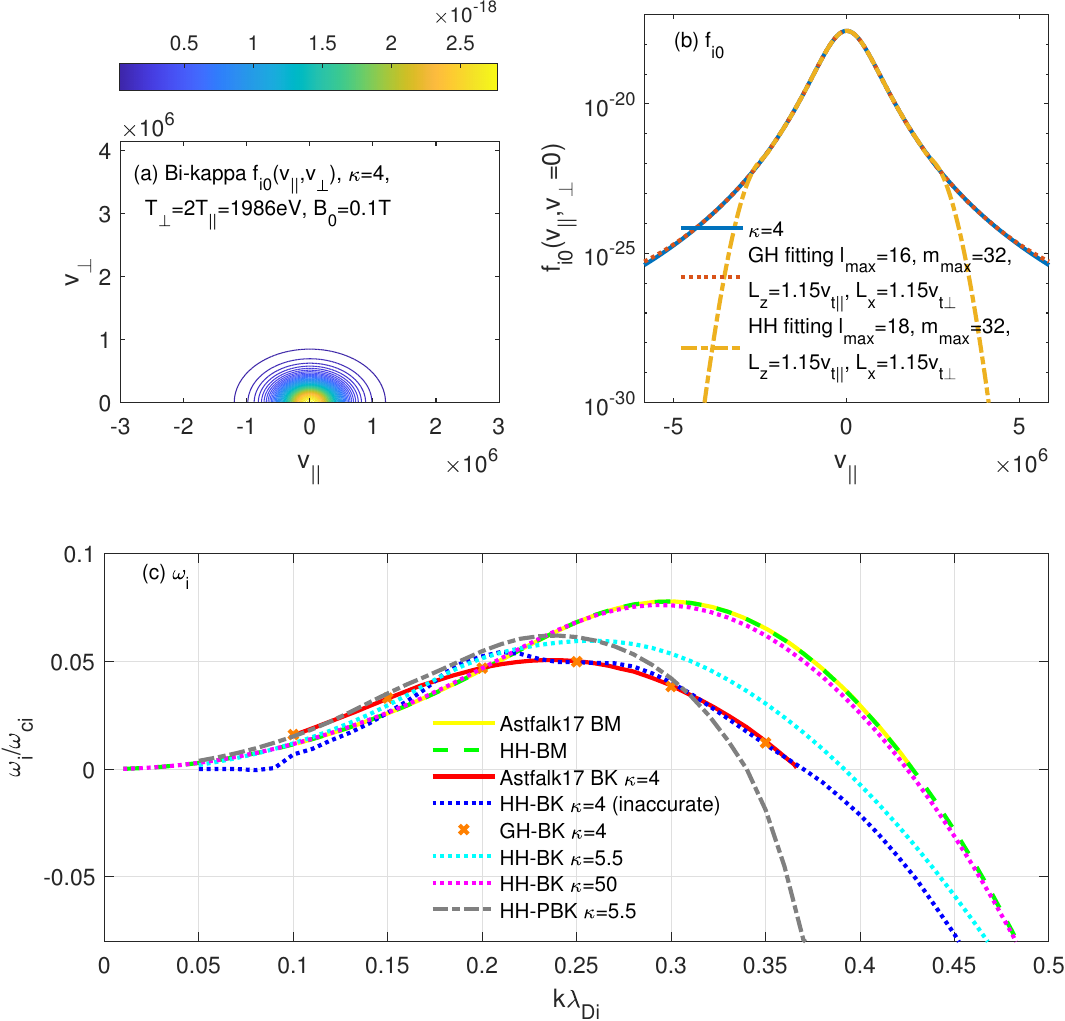}\\
\caption{Firehose instability for bi-Maxwellian (BM), bi-kappa (BK), and product bi-kappa (PBK) distributions at $\theta=45^\circ$.}\label{fig:bo_Astfalk17_firehose_theta=45}
\end{figure}

\begin{figure}
\centering
\includegraphics[width=8.5cm]{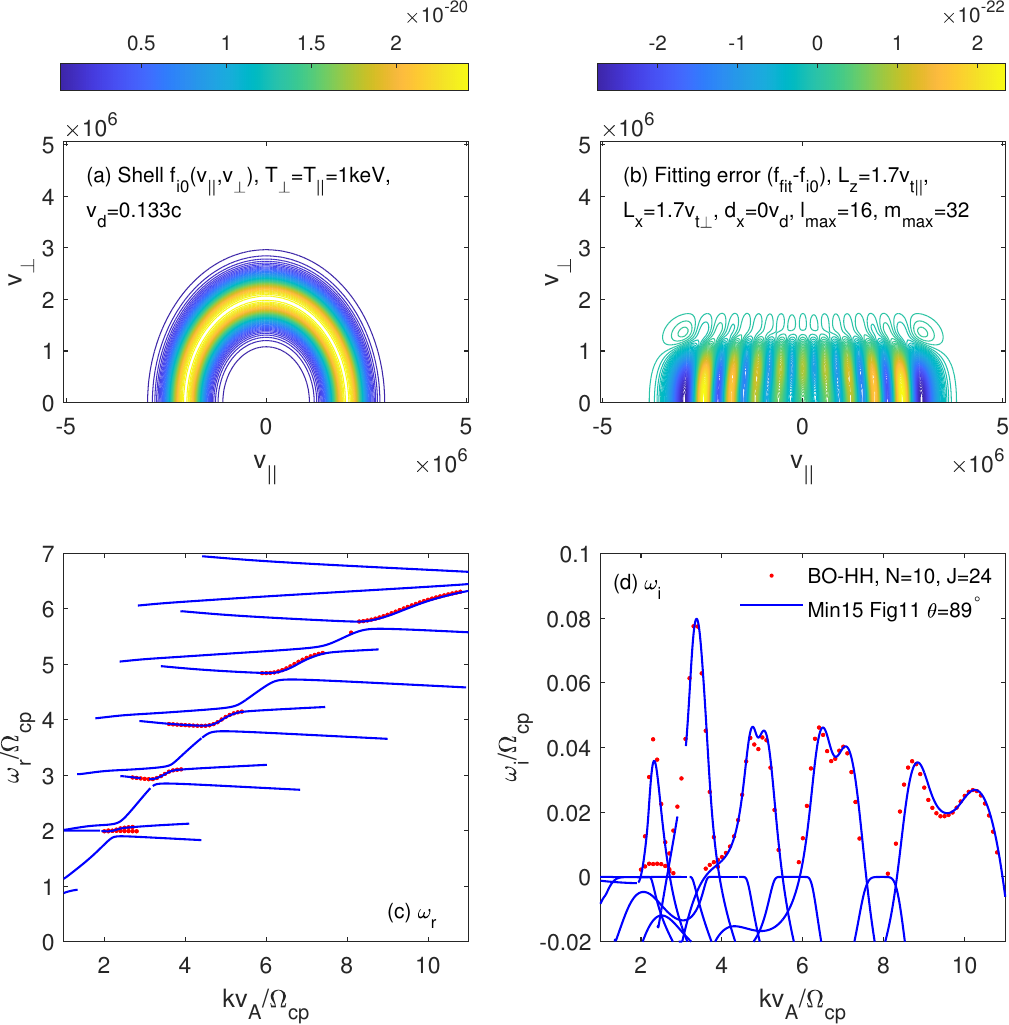}\\
\caption{Instabilities for shell ion distributions at $\theta=89^\circ$. The total CPU time to compute all 100 wave vector points is approximately 70 minutes for $S=3$, $J=24$, and $N=10$.}\label{fig:bo_Min15_shell}
\end{figure}

We apply the above framework to solve several special examples to demonstrate its performance. The first example involves the ring-beam electron instabilities, using parameters from Fig.~1 of Ref.~\cite{Xie2019}. The results for $\theta = 40^\circ$ are shown in Fig.~\ref{fig:bo_Umeda12_ringbeam_theta=40}. These results agree well with the Maxwellian-based solvers~\cite{Xie2019, Min2015} and show that different expansion parameters, such as $L_{sz}$, $L_{sx}$, $d_z$, and $d_x$, can yield consistent solutions, even for damped modes (excluding strongly damped modes). For all 120 wave vector points, the total computation time is less than one minute to obtain all solutions (only the three unstable solutions are plotted here).
The second example, shown in Fig.~\ref{fig:bo_Irvine18_ICE}, uses parameters from Fig.~3.8 in Ref.~\cite{Irvine2018} to study ion cyclotron emission (ICE) driven by a ring ion beam distribution in a magnetized fusion device. The results show good agreement with Ref.~\cite{Irvine2018}. Our approach calculates all solutions in one step, ensuring no solutions are missed. This is a significant advantage over conventional solvers, which require testing different initial guesses to obtain solutions one-by-one.

Superthermal distributions are commonly observed in space plasmas and in ICRF heating in laboratory plasmas. Thus, we present a third example using the kappa distribution. The parameters are taken from Fig.~2 in Ref.~\cite{Astfalk2017} for the firehose instability, with results shown in Fig.~\ref{fig:bo_Astfalk17_firehose_theta=45}. We find that the HH expansion is accurate for $\kappa \geq 5$ and can handle non-integer values (e.g., $\kappa = 5.5$) as well as very large values (e.g., $\kappa = 50$) with $l_{max} = 16$. For $\kappa = 4$, the GH expansion yields good agreement with Ref.~\cite{Astfalk2017}. Results for the product bi-kappa distribution are also shown. The CPU time for each calculation is less than one minute. Notably, our approach supports damped solutions naturally, which are not supported by the solver in Ref.~\cite{Astfalk2017}. 
A more complex example is the shell distribution, as shown in Fig.~\ref{fig:bo_Min15_shell}, with parameters taken from Fig.~11 of Ref.~\cite{Min2015}. The results agree well with Ref.~\cite{Min2015}. Slight differences may arise due to the limited significant digits provided for input parameters in Ref.~\cite{Min2015} or differences in the accuracy of the expansion.

In summary, we demonstrate that the new framework performs well for a wide range of nearly arbitrary distributions, covering all frequency ranges and wave vectors. We also find that the fitting accuracy of the distribution function significantly affects the results\cite{Astfalk2017,Verscharen2018}, highlighting the importance of precise distribution function data for obtaining accurate solutions.
The most significant innovation of this work is the development of a method that does not require initial guesses, which simplifies root-finding for nearly arbitrary distributions. It should also be noted that achieving high-accuracy calculations for non-smooth distributions may require more than double precision due to round-off errors in high-order coefficients~\cite{Bilato2012}. 
This framework is particularly useful for studying instabilities. While we have tested only a limited number of cases for damped modes, it performs well for weakly damped modes but less effectively for strongly damped modes. This limitation arises from the analytical continuation from real functions $f_{s0}(v_\parallel, v_\perp)$ to fitting functions of complex $v_\parallel$, which are sensitive to the fitting parameters.
The present approach can be seen as a major extension of the Maxwellian distribution-based KDR solver BO~\cite{Xie2019} to handle arbitrary distributions. Further extensions to relativistic and nonuniform plasmas are also anticipated. 

The source code for this work is available at: https://github.com/hsxie/boarbitrary.

{\it Acknowledgments} We would like to thank Kyungguk Min for providing the benchmark data used in Figs.~\ref{fig:bo_Umeda12_ringbeam_theta=40} and \ref{fig:bo_Min15_shell}.

\onecolumngrid
\clearpage
\foreach \x in {1,2,...,22}{%
    \thispagestyle{empty}%
    \centering
    \includegraphics[page=\x, width=\textwidth]{boall_derivation.pdf}%
    \clearpage
}

\end{document}